\documentclass[oldversion]{aa}
\usepackage{graphicx}
\usepackage{txfonts}
\usepackage{amssymb}
\usepackage{amsfonts}
\usepackage{subfigure}
\usepackage{natbib}
\bibpunct{(}{)}{;}{a}{}{,}

\begin{document}

\title{Mid-infrared laser light nulling experiment using single-mode conductive waveguides}

\author{L. Labadie\inst{1,2},
        E. Le Coarer\inst{2},
	R. Maurand\inst{2},
        P. Labeye\inst{3},
	P. Kern\inst{2},
	B. Arezki\inst{2},
	J.-E. Broquin\inst{4}}

\offprints{{}L. Labadie\\
           \email{labadie@mpia.de}}

\institute{Max Planck Institut f\"ur Astronomie, K\"onigstuhl 17, 69117 Heidelberg, Germany
	   \and
           Laboratoire d'Astrophysique de l'Observatoire de Grenoble, BP 53, 38041 Grenoble Cedex 9, France
           \and
           Laboratoire d'Electronique et des Technologies de l'Information (LETI), CEA-DRT-LETI, 17 rue des Martyrs, 38054 Grenoble Cedex 9, France
	   \and
	   Institut de Micro\'electronique, Electromagn\'etisme et Photonique, 23 rue des Martyrs, BP257, 38016 Grenoble Cedex 1, France}

\date{Received 2 January 2007; 25 April 2007}

\abstract{{\it Aims}: In the context of space interferometry missions devoted to the search of exo-Earths, this paper investigates the capabilities of new single mode conductive waveguides at providing modal filtering in an infrared and monochromatic nulling experiment; {\it Methods}: A Michelson laser interferometer with a co-axial beam combination scheme at 10.6 $\mu$m is used. After introducing a $\pi$ phase shift using a translating mirror, dynamic and static measurements of the nulling ratio are performed in the two cases where modal filtering is implemented and suppressed. No additional active control of the wavefront errors is involved. {\it Results}: We achieve on average a statistical nulling ratio of 2.5$\times$10$^{-4}$ with a 1-$\sigma$ upper limit of 6$\times$10$^{-4}$, while a best null of 5.6$\times$10$^{-5}$ is obtained in static mode. At the moment, the impact of external vibrations limits our ability to maintain the null to 10 to 20 seconds.; {\it Conclusions}: A positive effect of SM conductive waveguide on modal filtering has been observed in this study. Further improvement of the null should be possible with proper mechanical isolation of the setup.
}
\authorrunning{L. Labadie et al.}
\titlerunning{Mid-IR nulling experiment using single-mode conductive waveguides}

\maketitle

\section{Introduction}\label{intro}

The study of extrasolar planetary systems has become a growing field of modern astronomy in the last ten to fifteen years. After the discovery of an exo-Jupiter in close orbit around 51 Peg by \citet{Mayor}, the number of detected exoplanets has greatly increased, thanks to improved astronomical methods in terms of accuracy and sensitivity. A large proportion of exoplanets have been discovered by mean of indirect methods such as radial velocities (RV), but the direct detection -- or even direct imaging  -- of planetary companions constitutes a further step in our understanding of the physics of these objects. The case of exo-earths around Solar-type stars is of particular interest. The detection of these low mass planets is particularly challenging because of the tiny effects they induce on their parent star. 
Direct observation of Earth-like planets is limited by extreme contrast and small angular separation existing between the two bodies. Probing the habitable zone of such solar systems would face flux contrast of about 10$^{\rm 6}$ --10$^{\rm 7}$ in the mid-infrared with angular separations below 100 mas \citep{Angel86}.\\
Space-based missions like Darwin in Europe \citep{Fridlund04} or TPF-I in the United States \citep{Beichman} aim to detect and characterize Earth-size planets using mid-infrared nulling interferometry. 
The technique is based on starlight cancellation by means of destructive interferences in the mid-infrared, i.e from 5 $\mu$m to 20 $\mu$m \citep{Bracewell,Leger96,Angel97}. This spectral band presents the advantage of a lower star--companion contrast by three order of magnitude with respect to the visible domain \citep{Bracewell2}.\\
Nulling interferometry faces several instrumental obstacles that contribute to degrade the interferometric null. These include fine path-length control, intensity balance and polarization matching between the incoming beams, the requirement of a broadband achromatic $\pi$ phase shifter \citep{Labeque04} and fine control of the wavefront errors. As a consequence, the experimental demonstration of a deep broadband nulling has been actively pursued during the last years, first in the visible range \citep{Serabyn99,Wallace00,Haguenauer06} then at 10 $\mu$m \citep{Wallace05,Gappinger05}, in order to validate critical technologies. Among the different mentionned instrumental issues, the control of the wavefront corrugations is of major importance since the null is greatly degraded either by low-order errors ( i.e. residual optical path difference (OPD), pointing errors, optical aberrations) or high-order errors (spatial high-frequency defects due to optics and coatings) \citep{Leger95}. A reduction of wavefront errors is achievable using singlemode waveguides \citep{Shaklan88}, a solution commonly used in optical and near-infrared interferometry with optical fibers or integrated optics. \citet{Mennesson} have also underlined theoretically the advantages of singlemode waveguides for mid-infrared nulling interferometry to relax the strong instrumental constraints set on a deep nulling ratio. In a previous paper \citep{LabadieAA}, we presented the work achieved in the context of research and development on mid-infrared singlemode guided optics for stellar interferometry. The original concept was based on the manufacturing of hollow metallic waveguides (HMW) and their characterization in the laboratory via the analysis of the polarization analysis of the transmitted flux. That paper did not include results on nulling extinction ratios. As a futher step, conductive waveguides, for which application to nulling interferometry has been theoretically investigated by \citet{Wehmeier}, have been used in the present study to explore their impact as modal filters in a monochromatic light nulling experiment.\\
The present paper is organized as follows. First, we describe the goals of the experiment  and the made assumptions. Then, we describe the experimental setup and the adopted protocols. The final section presents the results on dynamic and static measurements.

\section{Strategy for the null measurement}

\subsection{Dynamic and static null}\label{strategy1}

The primary goal of this study is to obtain a direct comparison between the monochromatic extinction ratio that can be achieved with and without using a single-mode HMW, and to demonstrate that the effect of the residual wavefront errors on the interferometric null can be minimized using such a component. To obtain a valid comparison of the two mentionned cases (with and without waveguide), some precautions need to be taken on the physical meaning of the measured quantities. 
If we consider a monochromatic interferometer using the co-axial recombination scheme and fed by an unresolved source, the interferometric intensity $I$ measured on the single detector is given by

\begin{eqnarray}
I&=&I_{1}+I_{2}+2\sqrt{I_{1}I_{2}}V.\cos(\phi)\label{fringeseq}
\end{eqnarray}

\noindent $I_{1}$ and $I_{2}$ are the intensities in each arm
and $\phi$ the phase difference due to OPD. $V$ is the visibility term, with 0$<$$V$$<$1, due to low-order and high-order errors of the recombined wavefronts. The monochromatic rejection ratio is extracted from the visibility term $V$ through 

\begin{eqnarray}
\rho&=&\frac{1+V}{1-V}\label{nullterm}
\end{eqnarray}

\noindent In a real nulling experiment, $V$ is also affected by the spectral bandwith of the emission line if a laser source is used. This can have some impact if a deep nulling ratio is expected, but in the present context this issue can be neglected as explained in Sect.~\ref{tuning}.\\
\\
Following the preliminary results of \citet{LabadiePhD}, we propose to perform the nulling measurement in two different ways: a {\it dynamic} mode and a {\it static} mode. In the first approach (i.e. {\it dynamic}), a large number of fringe patterns containing two periods is recorded with a good sampling of the constructive and destructive fringe. An {\it a posteriori} measurement of the average photometric values $<$$I_{1}$$>$ and $<$$I_{2}$$>$ is performed to correct for the flux unbalance. Then, the {\it statistical} value $V_{i}\cos(\phi)$=($I_{i}$-$<$$I_{\rm 1}$$>$-$<$$I_{\rm 2}$$>$)/(2$\sqrt{<I_{\rm 1}><I_{\rm 2}>}$) for the $i^{th}$ occurrence is computed. An average value $<$$V$$>$ of the visibility is computed using the well-known statistical mean estimator $\bar{\mu}$ given by (1/$n$).$\sum_{i=1}^n V_{i}$, where $n$ is the number of samples \citep{Protassov}. The error bar on the mean visibilty is computed with the standard deviation estimator $\bar{\sigma}$ = $\sqrt{(1/(n-1))\sum_{i=1}^n (x_{i}-\bar{\mu})^2}$. The advantage of the dynamic method is that the error bar on the visibility is not obtained by computing the error propagation from $<$$I_{1}$$>$ and  $<$$I_{2}$$>$ -- this would be the case if only one occurrence of the fringe pattern was recorded -- but from the variance $\sigma^{2}_{V}$ of the sample $V_{i}$ which includes the uncertainty on the photometric channel. The second approach (i.e. {\it static}) involves manually optimizing by iterative process the destructive state and recording the transmitted signal over a significant laps of time. The nulled signal is compared to the constructive output obtained with the same principle. Here, the intensity unbalance is minimized {\it before} nulling the signal, either by partially masking the brightest channel with a small screen translated into the beam, or by playing with the tilt of this same channel in order to degrade the corresponding coupling efficiency (see Sect.~\ref{tiltimpact}). The advantage of the {\it static} method is that it can be used to obtain information on the temporal stability of the null. 



\subsection{Impact of the tilt on the photometric calibration}\label{tiltimpact}


\begin{figure}[t]
\centering
\includegraphics[width=8.5cm]{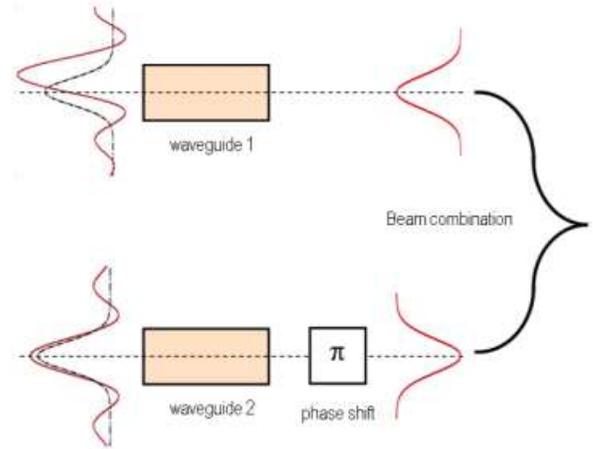}
\caption{Schematic view of modal filtering of the interferometric beams {\it before} recombination. By tilting the wavefront in one channel, the coupling efficiency is degraded. This impacts only the amplitude of the fundamental mode, not its phase.  }\label{combination1}
\end{figure}

Let us suppose that the incoming beams are coupled into two identical singlemode waveguides {\it before} beam recombination as it occurs, for instance, in an integrated optics beam combiner. The ratio of power coupled into each waveguide is given by the overlap integral 
between the electric field of the beam focused on the waveguide input and the electric field of the fundamental mode. For the sake of simplicity, we consider here only the one-dimensional case. We also consider identical linear polarizations for the overlapped fields, so that the vector product simply becomes a scalar product. Let us model the fundamental mode by a field distribution $S(x)$ 
and the excitation field of the focused beam by $E(x)$, 
where $x$ is the linear coordinate in the focal plane. For the purpose of numerical simulations, $S(x)$ and $E(x)$ are often modeled by, respectively, a Gaussian and a sine cardinal functions \citep{Jeunhomme}.
Under the previous conditions, the power guided by the fundamental mode of a singlemode waveguide is given by

\begin{eqnarray}
P(x_{\alpha}) & = &\frac{\left|\int_{\infty} E(x-x_{\alpha})\cdot S(x)\cdot dx\right|^{2}}{\int_{\infty} \left|S(x)\right|^{2}dx}\label{couplingfactor}
\end{eqnarray}

\noindent The term $x_{\alpha}$ = $\alpha$$\cdot$$f$ is the linear displacement in the focal plane of the focusing optics with respect to the geometric center of the waveguide. The value of $x_{\alpha}$ is linked to the relative tilt of the wavefronts $\alpha$. The integrals are computed over an infinite section transverse to the propagation axis. 
For two unbalanced channels $P_{\rm 1}$ and $P_{\rm 2}$, the tilt on the brightest beam induces 
a linear translation in the focal plane, which modifies the coupling efficiency. 
The term $P(x_{\alpha})$ decreases 
but the wavefront phase inside the waveguide remains unchanged due to the fundamental property of the singlemode waveguide.
The principle is illustrated in Fig.~\ref{combination1}.\\
At the time of this study, we did not have identical singlemode waveguides to filter each beam separately. Thus, we implemented modal filtering {\it after} beam recombination, which has the advantage to correct any additional phase error 
induced by the beam splitter. 
Also, even if no wavefront corrugation occurs after recombination, two slightly different properties of the waveguides (dimensions, metallic coating...) would introduce a differential effect on the filtered beams that will degrade the null.\\
The principle of modal filtering remains unchanged when it is implemented {\it before} or {\it after} beam combination. From Eq.~\ref{couplingfactor}, the integrated {\it amplitude} coupled to the fundamental mode can be written as

\begin{eqnarray}
A(x_{\alpha}) & = &\frac{\int_{\infty} E(x-x_{\alpha}).S(x).dx}{\left(\int_{\infty}\left|S(x)\right|^{2}dx\right)^{1/2}}
\end{eqnarray}

\noindent where $A$ is a complex number taking into account the phase shift $\phi_{c}$ between $E$ and $S$. When modal filtering is implemented {\it after} beam combination, the power coupled to the fundamental mode from two $\pi$ phase-shifted input fields $E_{1}$ and $E_{2}$ is

\begin{eqnarray}
P_{0}(x_{\alpha}) & = &\frac{\left|\int_{\infty} \left[E_{1}(x)-E_{2}(x-x_{\alpha})\right]S(x).dx\right|^{2}}{\int_{\infty}\left|S(x)\right|^{2}dx}\label{before}
\end{eqnarray}


\noindent If Eq.~\ref{before} is rewritten as

\begin{eqnarray}
P_{0}(x_{\alpha})  & = & \left|A_{1}(0)-A_{2}(x_{\alpha})\right|^{2}\label{after}
\end{eqnarray}

\noindent then the equation corresponds to the case where the input fields $E_{1}$ and $E_{2}$ are coupled separately to the fundamental mode of the singlemode waveguide prior to recombination. This results from the fact that modal filtering is, just like beam recombination, a linear process with respect to field amplitudes. Thus the two operations are commutable. Such a property is used to minimize the intensity unbalance in the same way as shown in Fig.~\ref{combination1}.

\section{Experiment description}\label{expdescription}

\subsection{The laboratory setup}\label{setup}

\begin{figure}[t]
\centering
\includegraphics[width=8.5cm]{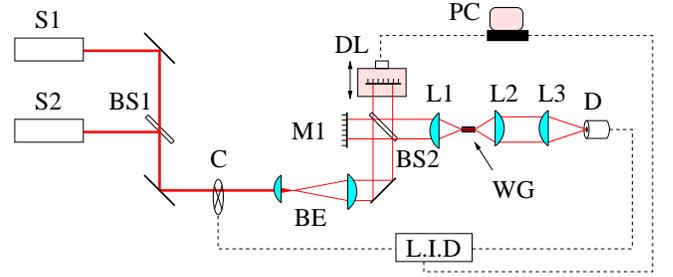}
\caption{Layout of the monochromatic interferometric bench. $S_{\rm 1}$: infrared CO$_{\rm 2}$ laser source; $S_{\rm 2}$: visible HeNe alignment source; $BS_{\rm 1}$: ZnSe beamsplitter; $C$: chopper; $BE$: beam expander; $BS_{2}$: ZnSe beam combiner; $DL$: delay line; $M_{1}$: interferometer fixed mirror; $L_{\rm 1}$: $f$/1 aspheric injection lens; $L_{\rm 2}$+$L_{\rm 3}$: afocal imaging system; $WG$: waveguide sample; $D$: HgCdTe detector; $L.I.D$: lock-in detection; $PC$: computer for data processing. Dashed lines represent electric wires.}\label{layout}
\end{figure}

\begin{figure*}
\begin{minipage}{\textwidth}
\centering
\subfigure[]{\includegraphics[width=5.5cm, angle=0]{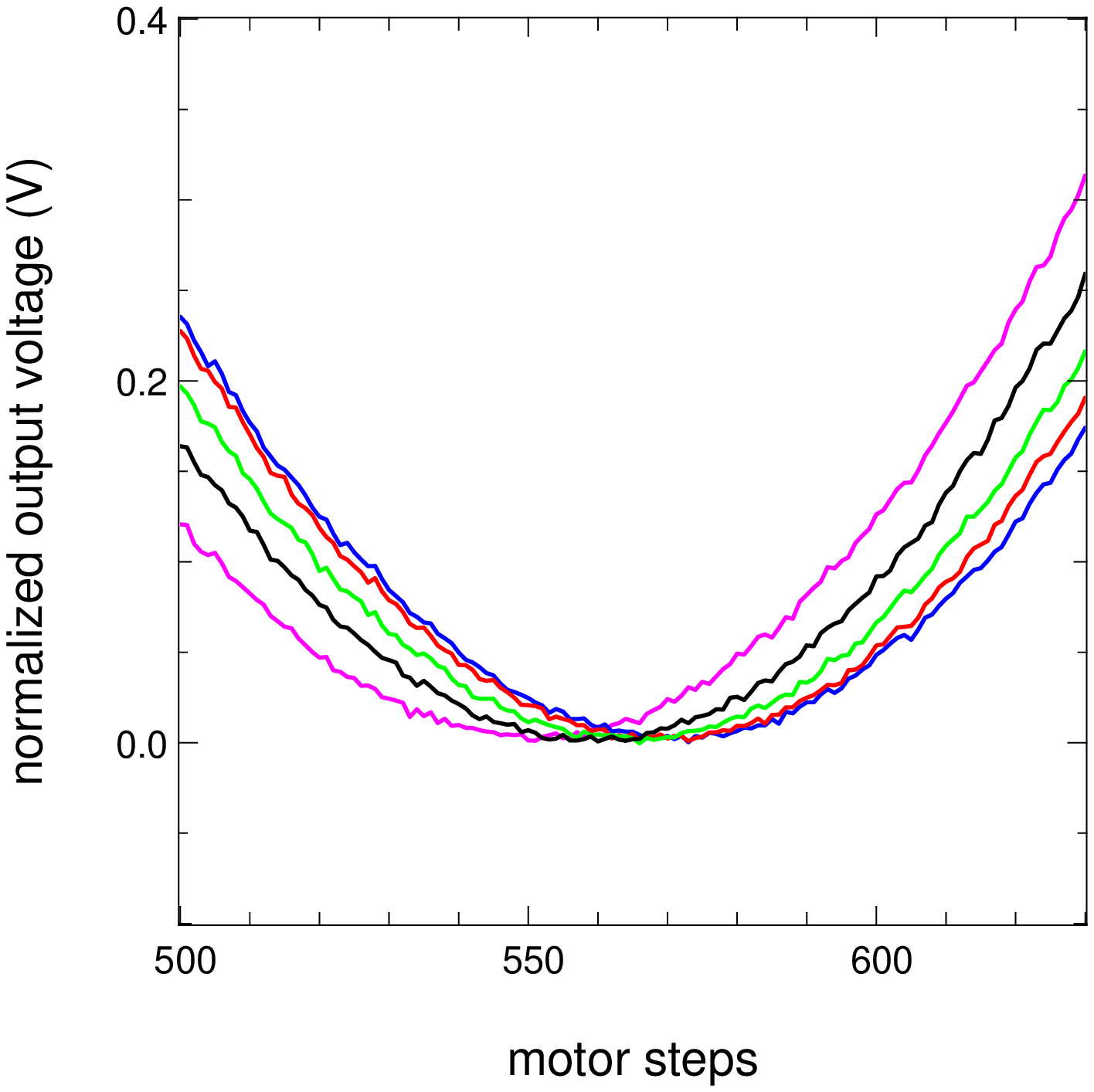}\label{fringe1}}
\hspace{0.5cm}
\subfigure[]{\includegraphics[width=5.7cm, angle=0]{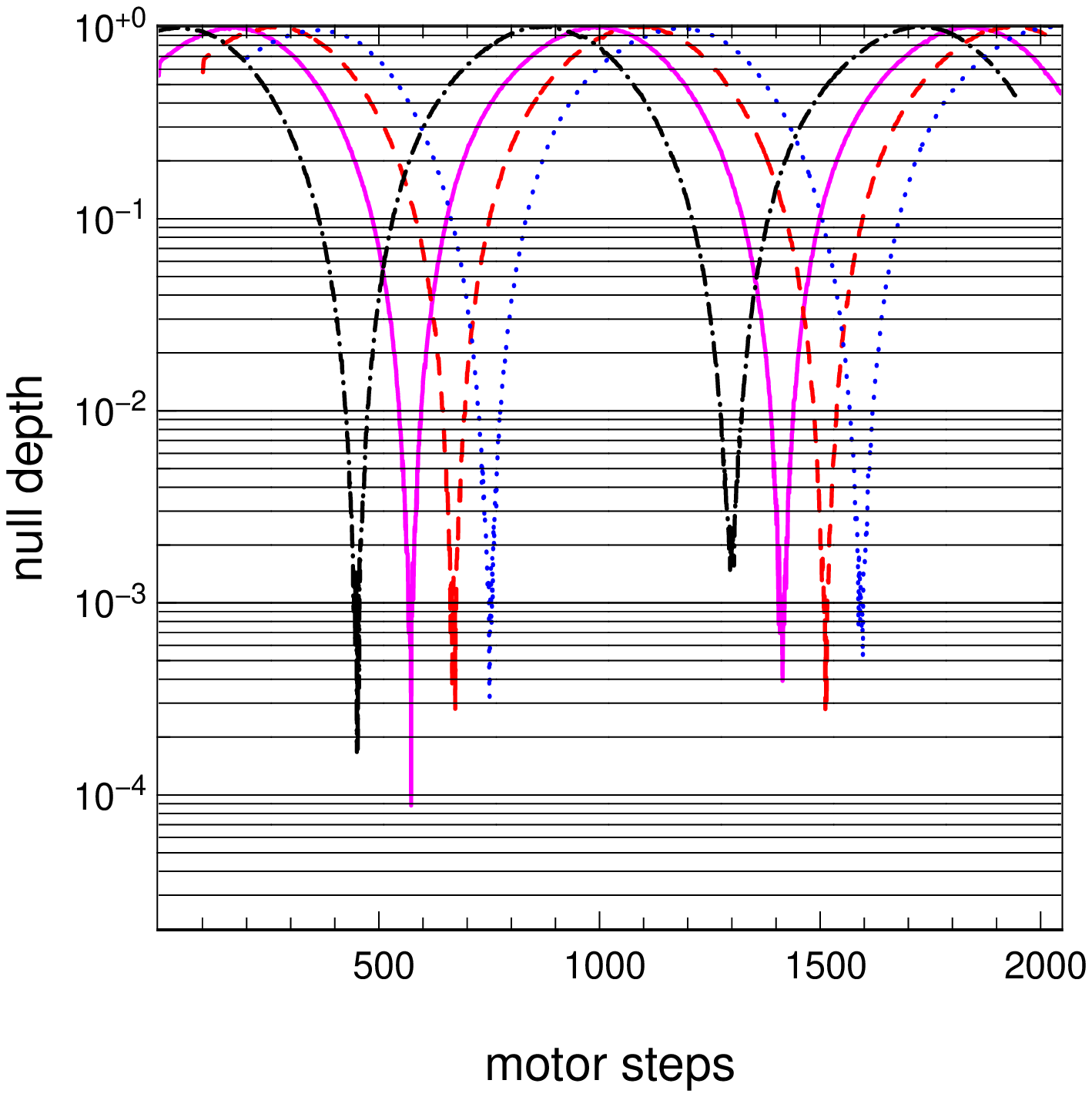}\label{nulldepth}}
\hspace{0.5cm}
\subfigure[]{\includegraphics[width=5.43cm, angle=0]{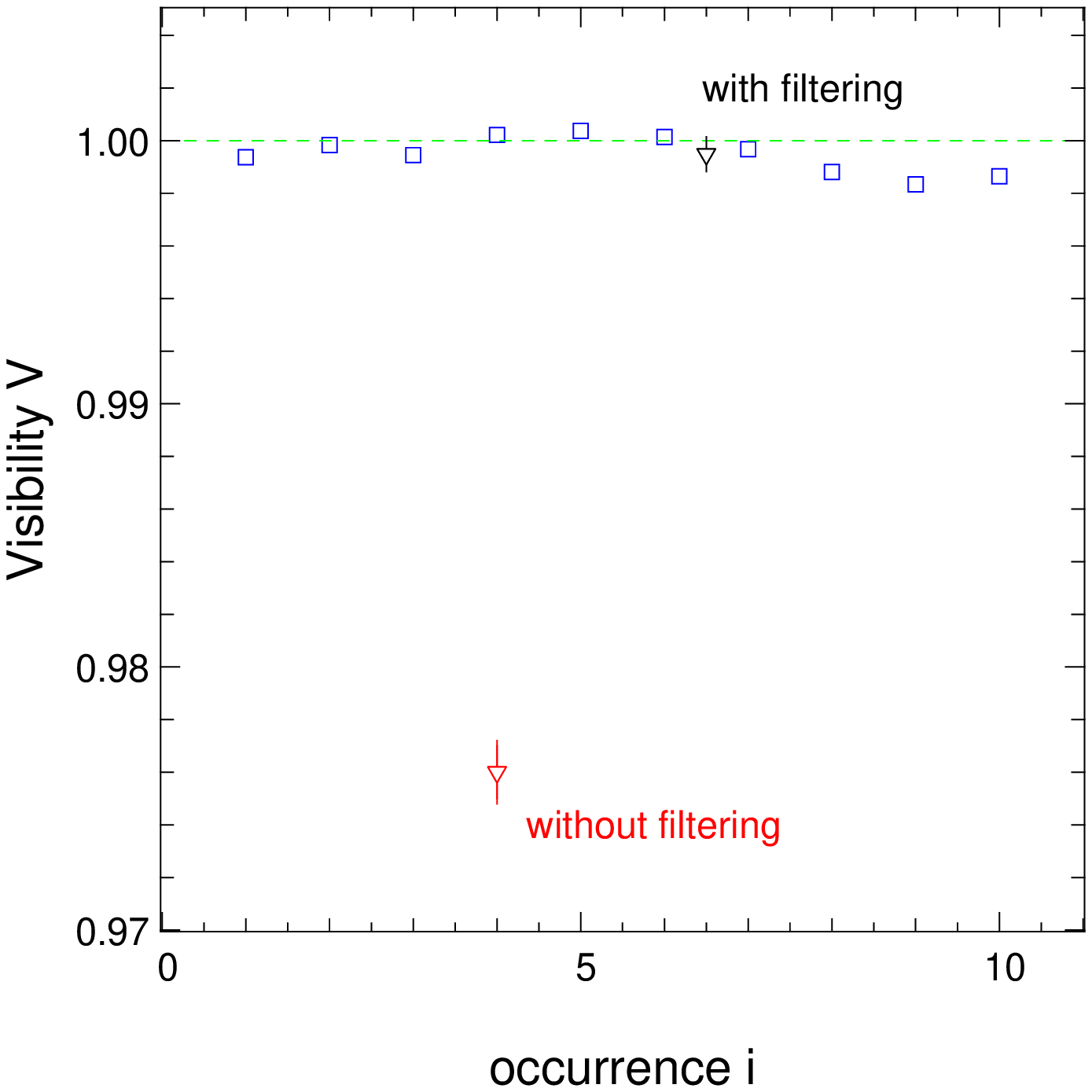}\label{fringe3}}
\end{minipage}
\caption{(a): Destructive fringes corrected from intensity unbalance and detector offset for five different occurrences of the fringe pattern. The position of the null remains unchanged within $\sim$500 nm of the delay line. One motor step is 12 nm. (b): Null depth as a function of delay line position in logarithmic scale with a modal filter. The four curves have been shifted horizontally  for better visibility. The curves show a null depth of few times 10$^{-4}$ and a maximum null of 7.7$\times$10$^{-5}$ is observed for the magenta curve (solid line). (c): Plot of the visibility for ten scans in dynamic mode. The two points with error bars give the mean visibility with and without a single mode HMW after beam recombination.}\label{triple}
\end{figure*}

The different elements of the setup were purchased in 2004 to first mount the injection bench used for the study in \citet{LabadieAA}. The setup was then adapted for nulling measurements during year 2005. The results presented in this paper are in the continuation of the preliminary results obtained in \citet{LabadiePhD}.\\
Our experiment consists in implementing a pupil-plane combination scheme in a classical Michelson interferometer. The layout of the testbench, described in Fig.~\ref{layout}, is adapted from the injection setup presented in \citet{LabadieAA}. The chip containing the singlemode waveguide {\it WG} used as a modal filter is mounted on a three axis positioner which allows precise placement of the sample input at the focal point of $L_{\rm 1}$. The hollow metallic waveguide is 1-mm long. Since in a conductive waveguide, the electric field is totally confined within the metallic cavity (the field is null in the metallic walls)
, the fundamental mode has the same size as the geometrical aperture of the waveguide, i.e. $\sim$10 $\mu$m in our case \citep{LabadieAA}. As a consequence, fast optics with f/1 or smaller is used for $L_{\rm 1}$ and $L_{\rm 2}$ to couple -- and decouple -- light for waveguides with high numerical aperture. The infrared source is a CO$_{2}$ laser emitting at 10.6 $\mu$m which is co-aligned with a 0.632 $\mu$m He-Ne laser. The temperature controller of the source helps to lock the laser on a given emission line. The $P_{\rm 22}$ line lasing at 10.611 $\mu$m with a spectral bandwidth $\Delta \nu$ $\approx$ 500 MHz was used in this study. A set of calibrated densities (not represented in the schematic view of Fig.~\ref{layout}) is placed in the optical train to attenuate the infrared laser, avoiding any damaging to the sample or the detection stage. The beam is reshaped to a diameter of 25 mm thanks to the beam expander {\it BE} which has a magnification {\it m} $\approx$ 7.\\
\noindent A zinc selenide (ZnSe) beam splitter {\it BS} is used in a double-pass scheme with the flat mirrors M$_{\rm 1}$ and M$_{\rm 2}$ to separate and recombine the wavefronts. The 8' wedge between the two faces of the beam splitter prevents interference from multiple reflected beams. In addition, an anti-reflection coating is applied to the rear face of {\it BS}, making the front face the reference plane for beam splitting. 
The two mirrors $M_{\rm 1}$ and $DL$ with a diameter of 25-mm are placed in tip-tilt mounts. In addition, $DL$ is translated thanks to a motor with piezo actuator to provide the delay line.\\ 
After $L_{\rm 1}$ couples the light into the waveguide, the output is re-imaged onto a 77K HgCdTe single-pixel detector $D$ with the afocal system composed of the plano-convex lenses L$_{\rm 2}$ and L$_{\rm 3}$. The f/2 numerical aperture of $L_{3}$ produces a 50-$\mu$m point spread function (PSF) that completely fits into the 500-$\mu$m square chip of the detector. 
Because the electronics of the HgCdTe detector is insensitive to the continuous component of the signal, the laser is chopped at the specific frequency of 191 Hz to avoid any contamination from the AC main supply 50 Hz harmonics. The chopper reference and the detected signals are processed through a classical lock-in amplifier to filter out unmodulated background. Finally, an 18 bit analog-digital converter card in a PC records the extracted signal.
The delay line scans about two fringes, which are sampled over 2048 points. Each point represents an increment of 12 nm every 90 ms, corresponding to a scan of 190 seconds.
The piezo actuator is controlled to repetitively perform the same OPD of the delay line. As shown in Fig.~\ref{fringe1}, 
a small shift of the dark fringe is observed due to an uncertainty in the absolute re-positioning of the delay line.
But, since this remains below $\sim$500 nm, it has a negligible effect on the measurement.\\
The experiment takes place in an open-air non-cryogenic environment. In addition, no active control of the OPD or of the wavefronts relative tilt is implemented so far. Thus the measurements are sensitive to air turbulence, mechanical vibrations or electronics drifts.

\subsection{Tuning the interferometer}\label{tuning}

\begin{figure}[b]
\centering
\includegraphics[width=5.5cm, angle=0]{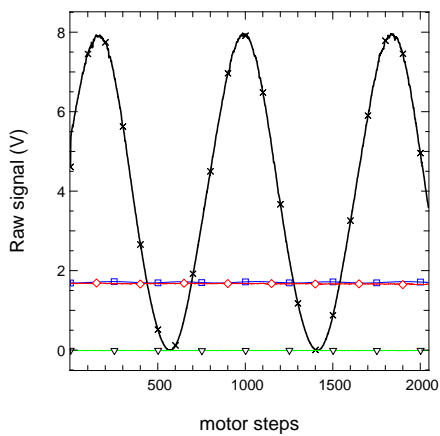}
\caption{Raw signals as a function of motor steps involved in the dynamic measurement of the interferometric null. The black curve (crosses) is the interferometric signal. The blue (squares) and magenta (diamonds) curves are the photometric channels. The green curve (triangles) is the detector offset. One scan has 2048 points recorded over 190 s. The linear step is 12 nm.}\label{rawdata}
\end{figure}

The interferometer is initially adjusted without any waveguide in the optical path. Using the tip-tilt mount of $M_{1}$ and $DL$, the two channels are first superimposed in constructive mode in the image plane of $L_{3}$ using a mid-infrared camera for the coarse alignment. The camera is then replaced with the HgCdTe detector connected to the lock-in amplifier.The OPD is then adjusted to reach a destructive state followed by a fine tuning of the wavefront tilt. This operation is made under conditions of destructive interferences rather than constructive ones because a tiny variation of the transmitted signal due to the relative tilt of the wavefronts is easily detectable with an almost nulled signal.  
It is not necessary to have perfect destructive interference to perform this tuning: a phase shift close to $\pi$ between the wavefronts is sufficient according to Eq.~\ref{fringeseq}.\\
Once the deepest achievable destructive signal is obtained from the tip-tilt adjustment, about ten scans of each photometric channel are recorded by successively masking $M_{1}$ and $DL$.
The setup is then returned to interferometric mode and about fifty scans of the fringe pattern are taken with the same geometrical OPD. 
The detector offset is recorded with the same procedure by simply turning off the laser sources. This measurement is made once before and once after the fringe acquisition. The plots of Fig.~\ref{rawdata} show a single occurrence of the different raw signals involved in the calibration of the null.\\
The same experimental procedure is followed when measuring the interferometric null using a singlemode waveguide in dynamic mode. The waveguide is simply introduced in the optical path using the three axis positioner and  its position optimized by maximizing the transmitted flux. The lens $L_{3}$ is then translated by 1 mm -- the geometrical length of a conductive waveguide -- to conjugate optically the waveguide output plane with the detector plane.\\
When using geometric translation with a mirror as a phase shifter, the interferometric null is wavelength dependent. Although the laser source has a very long coherence length, it cannot be considered as infinite in the frame of a co-axial nulling experiment. For a source with spectral width $\Delta \lambda$, the corresponding visibility loss is given by  {\it 1}-sinc($\pi\Delta z\Delta\lambda/\lambda^2$) where $\Delta z$ is the distance from the zero-path difference point. In this experiment, the interferometer is tuned to zero-OPD by measuring geometrically the position of the translating mirror with respect to the uncoated face of the beam splitter. The accuracy of this measurement is better than $\Delta$z = 1 mm, which corresponds to a visibility loss $<$ 2.2$\times$10$^{-6}$ for the $\sim$0.187 nm spectral width of the $P_{\rm 22}$ emission line of the laser.

\section{Experimental results}\label{results}

The graph of Fig.~\ref{nulldepth} presents the null depth for four different fringe dynamic acquisitions of our sample with modal filtering. Each data set is corrected for the detector offset and for the photometric unbalance using the same quantities $\bar I_{\rm 1}$, $\bar I_{\rm 2}$ for all the sets (see Sect.~\ref{strategy1}). The values $\bar I_{\rm 1}$=1.70568$\pm$0.008 V and $\bar I_{\rm 2}$=1.68804$\pm$0.008 V were recorded for the two photometric channels, which results in a photometric unbalance of only $\sim$1\%$\pm$0.01\% before correction. The different curves have been artificially shifted by 150 motor steps for a better visibility of the plot.
A given color represents one snapshot of the normalized fringe pattern after correction. The presented curves show a null depth of a few times 10$^{-4}$ on the left part of the graph between 1 and 1000 counts. In one case the null reaches 7.7$\times$10$^{-5}$ (magenta solid line) for an average rejection ratio of 12,990:1. On the right part of the graph (i.e. after motor step 1000), one can observe that the null depth is slightly degraded for all the four curves. A possible explanation is that the motor of the delay line presents a slight but systematic drift when leaving the zero-OPD postion (around count i=100), which was used for the initial alignment of the interferometer (see Sect.~\ref{tuning}). Although this hypothesis needs to be experimentally confirmed, it appears as a possible limitation for a setup like ours which has no active control of mechanical drifts.\\
In the graph of Fig.~\ref{fringe3}, we plot the dynamic measurements of ten scans of the interferometric visibility represented by the blue squares. The visibility $V$=1.0 is plotted as a green dashed line. The two single points represent the mean visibility and the corresponding error bar when modal filtering is implemented (black triangle) and when no filter is used in the optical path (red triangle). These two points are obtained through statistical measurements of the visibility in dynamic mode. Some points of the blue statistical serie appear above unit visibility, which would mean that the rejection ratio is infinite. However, this only corresponds to a bias in the substraction of the mean detector offset: at the instant $t$ where such a point is acquired, the instantaneous value of the offset was likely below the mean offset. The visibility curve in Fig.~\ref{fringe3} shows a correlation between one dynamic scan to the other. Its origin is at the moment unknown and this effect is included in the error bar of the mean visibilty. The average visibility is found to be $V$=0.999486$\pm$0.0007, which translates to a mean extinction ratio $\rho^{-1}$=2.5$\times$10$^{-4}$. We obtain good accuracy since the data  is not affected by laser intensity fluctuations in the coherent combination (the power drifts affect the two channels of the interferometer equally). The error bar overlaps the visibility $V$=1. This indicates that the achievable null might be better than the previous value, although ultimately limited by the dynamic range of the detector and lock-in system which reaches the 10$^{-6}$ level.\\
The second point -- placed arbitrarily at the fourth occurrence -- gives the average visibility and error bar of the dynamic measurement without modal filter in the optical path and with the same alignment settings. The degradation can be clearly observed with an average visibility of $V$=0.976$\pm$0.001. The corresponding extinction ratio is 1.2$\times$10$^{-2}$.\\
\\
\noindent The dynamic acquisition method is an interesting statistical approach to measuring the achievable extinction ratio, but because the destructive fringe is scanned, the method does not provide large information on the stability of the null with the current setup. Such information is obtained through a {\it static} measurement of the null. The principle is to search manually for the optimal destructive signal with the piezo motor and then to minimize as much as possible the intensity unbalance either by playing with the coupling of the strongest channel or by partially masking with a small screen that can be translated into the beam (see Sect.~\ref{strategy1}). Prior to any null measurement, the time stability was investigated to understand the potential impact of external constraints (vibrations, drifts...).\\
\begin{figure}[t]
\centering
\includegraphics[width=7.5cm, angle=0]{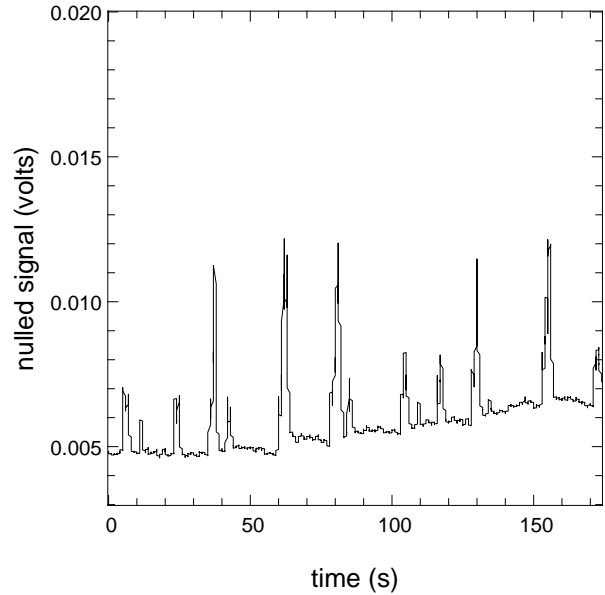}
\caption{Nulled output as a function of the time. This plot shows that the time scale over which the null can be considered stable is 10 to 20 seconds. The strong spikes come from sensitivity of the setup to vibrations and external constraints.}\label{stability}
\end{figure}
The plot of Fig.~\ref{stability} shows the raw uncalibrated null signal -- i.e. not subtracted from the detector offset -- recorded over 170 seconds. This plot does not correspond to the the best-null case measured later on. The plot shows that the nulled signal remains stable for a maximum of 10 to 20 seconds at this null level. Several spikes due to nulling degradation appear after this time scale. These spikes are very likely caused by vibrations from which the setup is not isolated at the moment. We also observe a long-term drift of the signal at the end of the plot suggesting that some positioning element in the setup might suffer from a constrain at low frequencies. This result implies that a destructive state can be steadily maintained in the best case for only ten seconds or so.\\
A static measurement of the null was performed with the two methods of photometric equalization described above. For the equalization by screen translation, the results are shown in Table~\ref{finalnull}.

\begin{table}[h]
\centering
\begin{tabular}{| l | c |}                         \hline 
Constructive signal & 620 $\pm$ 4 mV                          \\ 
Uncalibrated null   & 0.042 $\pm$ 0.005 mV                       \\
Detector offset     & 0.007 $\pm$ 0.003 mV                      \\ \hline 
Extinction ratio & 5.6$\times$10$^{-5}$ (17,820:1) \\ \hline
\end{tabular}\\
\caption{Experimental data obtained during null measurement in static mode.The extinction ratio is computed from the mean values of the voltage.}\label{finalnull}
\end{table}

\noindent The recorded null value was obtained over a period of 10 seconds with a good level of accuracy. On the other hand, adjusting the coupling efficiency to equalize the photometry did not provide better results than that presented in Fig.~\ref{triple}.\\
\\
To which extent can the {\it dynamic} and {\it static} approaches be compared? The plot in Fig.~\ref{stability} shows a null stability over a time scale of 10 to 20 seconds, which is shorter than the duration of a single scan in {\it dynamic} mode (190 s). However, within these 190 seconds, the interference is deeply destructive for approximatively 10 seconds (i.e. 100 motor steps, which corresponds to 9 s, see Fig.~\ref{rawdata}), while the constructive state is barely affected by the setup instabilities. The {\it static} approach aims at giving {\it one} occurrence (possibly the best one) of the {\it dynamic} mode, in order to confirm the potential of our modal filters. Since the {\it dynamic} mode cannot be guaranteed to fall on an optimal destructive state within one given scan, this aspect is reflected in the experimental dispersion of the corrected visibilities.\\
Because the {\it best case} nulling ratios obtained are comparable in both approaches (7.7$\times$10$^{-5}$ in {\it dynamic} mode, 5.6$\times$10$^{-5}$ in {\it static} mode), it could be initially inferred that the two methods are equivalent. This could be true at the experimental level, but not at the level of the instrument system.
The {\it static} approach would be clearly worthwhile in a vibration-free environment, although usually difficult to obtain, since this would permit to spend larger integration time on the dark fringe, limit the number of detector read outs and bleeding effects due to switching between dark and bright fringe.
At the contrary, the {\it dynamic} mode would appear more favorable when external constraints affect the experiment because: 1. The instrument stability could be quantified statistically through the standard deviation of the measured visibilities. 2. The scans corresponding to a deep null could be isolated in a deterministic way.\\
In addition, we think that the importance of the stability issue favors the use of compact and stable integrated optics beam combiners.

\section{Conclusions}\label{disc}

The experiment presented in this paper has shown that the use of a singlemode conductive waveguide as a modal filter permits significant improvement in the extinction ratio in a 10-$\mu$m nulling interferometer. The ratios were measured in a dynamic way by acquiring a large number of fringe patterns which gave a statistical value of the null. An average ratio of $\rho^{-1}$=2.5$\times$10$^{-4}$ has been measured with an error bar of $\pm$0.07\% on the visibility. A deeper null may have been possible in theory given the error bars, but our setup did not permit us to measure it in dynamic mode. A static measurement of the null has provided a single occurrence of 5.6$\times$10$^{-5}$. With the current setup, such a null can be maintained for approximately 10 to 20 seconds in the best case.\\
It is clear that in the frame of a mid-infrared nulling experiment, a proper isolation of the setup from external vibrations, electronic drifts of the delay line or variations of the local temperature are necessary. These types of issues are at the moment a limiting factor.\\
This study aims to show that singlemode conductive waveguides can be efficient modal filters over distance of 1-mm in the context of mid-infrared nulling interferometry. Theoretical studies on the filtering length of conductive waveguides for infrared radiation suggest that the waveguides could be even shorter to compensate for high propagation losses \citep{Tiberini07} without altering the filtering capabilities.

\begin{acknowledgements}
      This work was supported by the \emph{European Space Agency} ($\it
      {ESA}$) under contract 16847/02/NL/SFe and supported by
      funding from the \emph{French Space National Agency} ($\it
      {CNES}$) and {\it Alcatel Alenia Space}. The authors thank Dr. T. Herbst for fruitful comments.
\end{acknowledgements}

\bibliographystyle{aa}
\bibliography{paper_gc.bib}

\end{document}